\documentclass[floatfix,aps,twocolumn, superscript address] {revtex4-2}
\usepackage{hyperref}
\hypersetup{
colorlinks=true, citecolor=blue, urlcolor=blue, linkcolor=blue}
\usepackage{graphicx}
\usepackage{dcolumn}
\usepackage{color}
\usepackage{xcolor}
\usepackage{bm}
\usepackage{amsmath,amssymb}
\usepackage{enumerate}
\usepackage{array}
\usepackage{multirow}
\usepackage{gensymb}
\usepackage{comment}
\usepackage{booktabs}
\usepackage{soul}

\usepackage{pdfpages}
\makeatletter
\AtBeginDocument{\let\LS@rot\@undefined}
\makeatother

\begin {document}
\title{Gate voltage modulation of the superconducting state in a degenerate semiconductor}

\author{Bikash C. Barik}
\affiliation{Department of Physics, Indian Institute of Technology Bombay, India}
\author{Himadri Chakraborti}
\affiliation{Department of Physics, Indian Institute of Technology Bombay, India}
\affiliation{SPEC, CEA, CNRS, Universite Paris-Saclay, CEA Saclay, 91191 Gif sur Yvette Cedex France}
\author{Buddhadeb Pal}
\affiliation{Department of Physics, Indian Institute of Technology Bombay, India}
\affiliation{Indian Association for the Cultivation of Science, Kolkata, 700032, India}
\author{Aditya K. Jain}
\affiliation{Department of Physics, Indian Institute of Technology Bombay, India}
\affiliation{Department of Physics, Royal Holloway, University of London, Egham, Surrey, TW20 0EX, UK}
\author{Swagata Bhunia}
\affiliation{Department of Physics, Indian Institute of Technology Bombay, India}
\author{Sounak Samanta}
\affiliation{Department of Physics, Indian Institute of Technology Bombay, India}
\author{Apurba Laha}
\affiliation{Department of EE, Indian Institute of Technology Bombay, India, India}
\author{Suddhasatta Mahapatra}
\affiliation{Department of Physics, Indian Institute of Technology Bombay, India}
\author{K. Das Gupta}
\email{kdasgupta@phy.iitb.ac.in}
\affiliation{Department of Physics, Indian Institute of Technology Bombay, India}
\begin{abstract}


In this work, we demonstrate that the modulation of carrier density can alter the superconducting transition temperature by up to $204$ mK in epitaxial Indium Nitride on Gallium Nitride, accounting for the $10$\% of the transition temperature in ungated conditions. Our samples are likely free from strong localization effects and significant granularity, as indicated by \( k_f l \gg 1 \), suggesting that the primary determinant of the transition temperature in InN is carrier density, rather than disorder scattering. The observed behavior is consistent with BCS s-wave superconductivity, corroborated by the superconducting parameters we measured. Furthermore, we observed a $60$\% bipolar suppression of the supercurrent in our experiments.

\end{abstract}
\maketitle

The bare interaction between two particles in a many-body system is necessarily modified by the presence of the other particles and excitations. This modification or screening is a function of particle density, which in metallic systems is very difficult to vary.   Metals have carrier densities $\sim 10^{23}$ ${\rm cm^{-3}}$ and even a monolayer of metal will have an areal density of at least $\sim 10^{15}$ ${\rm cm^{-2}}$. These  can not be modulated  by any conventional solid-state gate in a Field Effect Transistor (FET) configuration. This limitation arises because achieving even $\sim$ 1 \% change in carrier density demands an unsustainable electric field. Even when utilizing very thin (thickness $d$) high-$K$ dielectrics, the breakdown field (approximately $\sim 1 \rm \ V/nm$) constrains the maximum achievable areal density modulation $\delta{n}\approx\left( \dfrac{{\epsilon_0}K}{e} \right)\left(\dfrac{V_G}{d}\right)$ to be on the order of $\sim 5\times 10^{13}{\rm \ cm^{-2}}$, where $V_G$ represents the gate voltage. In contrast, non-degenerate semiconductor channels offer the flexibility for density modulation due to their significantly lower bulk concentration, typically ranging from ($10^{16}$ to $10^{17}$ ${ \rm cm^{-3}}$).
Historically, this is the key reason superconducting pairing, ferromagnetic exchange etc. could not be gate-controlled because these usually occur in metals, alloys or oxides which have carrier densities much higher than $\sim$ $10^{20}$ $\rm cm^{-3}$. In recent years ionic liquid gating technique \cite{misra2007electric, daghero2012large, petach2014mechanism, ueno2011discovery, piatti2017control, yuan2010electrostatic,nakagawa2021gate} has brought ferro and antiferromagnetic interactions and potential superconducting interactions to gate tunable regime \cite{petach2014mechanism, walter2020voltage, ueno2008electric} to  a certain extent. It is interesting to ask whether an effective attractive interaction between electrons near the Fermi surface can persist in systems with significantly lower electron densities, characteristic of semiconductors (typically $10^{16}$ - $10^{20}$ $\rm cm^{-3}$).  Certain semiconductors, such as GeTe, B-doped diamond, B-doped Si and InN do show superconductivity. For instance, InN films exhibit a transition at temperatures around $T$ $\sim$ $2 - 3 \rm \ K$, heavily alloyed GeTe,  a transition is seen around  $T$ $<$ 0.3 K \cite{hein1964superconductivity, smith1977superconducting}. Diamond and Si, require carrier densities above $10^{21} \rm \ cm^{-3}$ for superconductivity \cite{ekimov2004superconductivity, bustarret2006superconductivity}. Interestingly, the carrier density required for superconductivity in InN is notably lower, falling within the range of ($10^{17}$ - $10^{18}$${ \rm \ cm^{-3}}$) \cite{miura1997anomalous, inushima2001anisotropic, inushima2006superconductivity},  which is  $\sim$ $10^{3}-10^{4}$ times less than conventional metals.

In this work, through controlled modulation of the carrier density by ionic liquid gating, we achieved a reversible change of the superconducting transition temperature ($T_C$) by $204 {\rm \ mK}$ with $\pm$ $3.25$ V, corresponding to approximately $10\%$ of the ungated $T_C$. In these devices, carrier densities were successfully changed by nearly $30\%$, and the critical current by more than $50\%$ compared to their ungated values. The areal capacitance between the gate and the channel was corroborated through independent methods, specifically, Hall effect measurements and electrochemical impedance spectroscopy (EIS) \cite{yuan2009high, yuan2010electrostatic}. It is important to note that previous investigations into the field effect on superconductors employed solid dielectric \cite{glover1960changes} and ferroelectric \cite{stadler1965changing} gates, yielding minimal effects. More recently, a shift of approximately $80{\rm \ mK}$ ($\Delta T_{C}$) was observed in Nb thin films subjected to applied gate voltages ranging from -$4$ V to $5$ V \cite{choi2014electrical}. On NbN thin films, a $\Delta T_{C}$ of around $85 {\rm \ mK}$ was achieved with applied gate voltages of $\pm 3$ V \cite{piatti2017control, piatti2016superconducting}. \\
Our prior experiments on angle-dependent critical field \cite{pal2018superconductivity} revealed that even in epitaxially grown $\rm{InN}$ films with a thickness of upto $300 {\rm \  nm}$, superconducting electrons are inherently confined to a surface layer with a thickness of $d_s \sim 25 {\rm \ nm}$—a dimension smaller than the coherence length $\xi_0$. Additionally, the significant density of surface states pins the surface Fermi level within the conduction band, eliminating the presence of a Schottky barrier at the surface and, concurrently, making hole-doping of the material exceptionally challenging \cite{jones2006evidence}.
However, it is precisely this combination of low bulk carrier density and the potential for a substantial response to perturbations in the surface states that renders ${\rm InN}$ an intriguing candidate for ionic-liquid gating where the strongest perturbation occurs close to the surface and gradually diminishes within a few Thomas-Fermi screening lengths as one moves into the bulk of the material.


\begin{figure}[ht]
	\centering
    \includegraphics[width=0.47\textwidth]{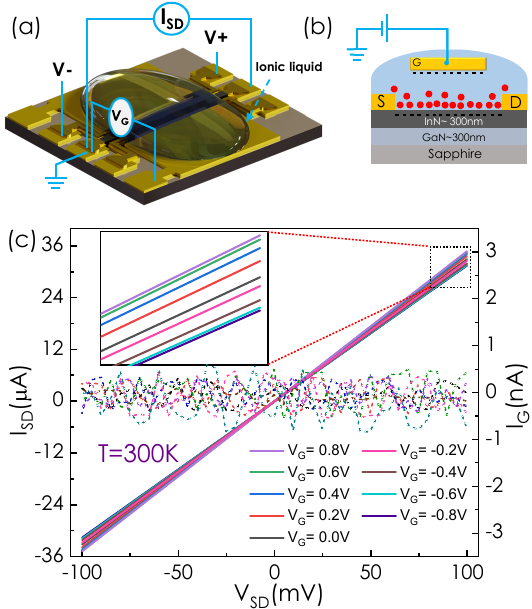}
    \caption{\textbf{Electric double-layer transistor \& Output characteristics}: (a) Schematic diagram of InN-IL gated transistor. The area of the gate electrode is much larger than the channel area. (see explanation in text). DEME-TFSI ionic liquid is used. (b) Cross sectional view. (c) Output characteristics at room temperature. (inset) shows a clear shift in current values with applied gate voltages. Gate leakages current are shown in the right panel. This data is taken from InN-B3601 sample.}
    \label{EDL_schematic}
\end{figure}

The InN samples were grown on sapphire substrates by plasma-assisted molecular beam epitaxy (PA-MBE) on c-oriented sapphire \cite{ghosh2017tuning, barick2016structural, barick2018network, pal2018superconductivity} with negligible oxygen impurity demonstrated by the absence of oxygen O1s core level transition in XPS \cite{ghosh2017tuning}. Patterned channels (Fig. \ref{EDL_schematic}(a)) on these were lithographically fabricated using $\rm Al_{2} O_{3}$ as a hard mask layer followed by chlorine-based dry etching and subsequent removal of the hard-mask by $\rm{HF}$. The devices had effective channel area of 1150 $\rm \mu m$ $\times$ 100 $\rm \mu m$ and thickness of $\sim$ 300 nm (Fig. \ref{EDL_schematic}(b)). Au/Ti (80 nm/20 nm) contacts were used for source (S) and drain (D) contacts.  The gate electrode is much larger in area than the channel, eliminating  the necessity of a reference electrode \cite{yuan2010electrostatic}. We used DEME-TFSI as the ionic liquid (IL). Contacts were covered with silicone gel during application of the IL. (See supplementary for detailed fabrication steps.). The ionic conductivity of DEME-TFSI are much higher than most other electrolytes \cite{misra2007electric}. The ionic liquid was cured in high vacuum at $\rm 110 \degree \ C $ for $\sim$ $48$ hrs.  All low-temperature measurements were carried out in a vacuum of $10^{-5}$ mbar or better. Fig. \ref{EDL_schematic}(c), shows typical output characteristics (i.e. $V_{SD}$ vs $ I_{SD}$ measurements) conducted at different gate voltages ($ V_{G}$), ranging from $+0.8$ V to $-0.8$ V with a step increment of $0.2$ V. The gate leakage remains $10^{3}$ times smaller than $ I_{SD}$. The observed trends and direct Hall voltage measurements at $300\rm K$, (as well as $4.8 \rm K$), confirm that the carrier density is stable and  indeed changes in a predictable way, allowing clear determination of the electron density ($n$) and mobility ($\mu_e$) from four terminal measurements.  We recorded a $ \sim 30\%$ change in carrier density with applied gate voltages, a finding of considerable relevance within the context of high carrier density systems \cite{petach2014mechanism, sagmeister2006electrically, ueno2011discovery, prassides2011superconductivity}. 

\begin{figure*}[t]
	\centering
	
    \includegraphics[width=0.85\textwidth]{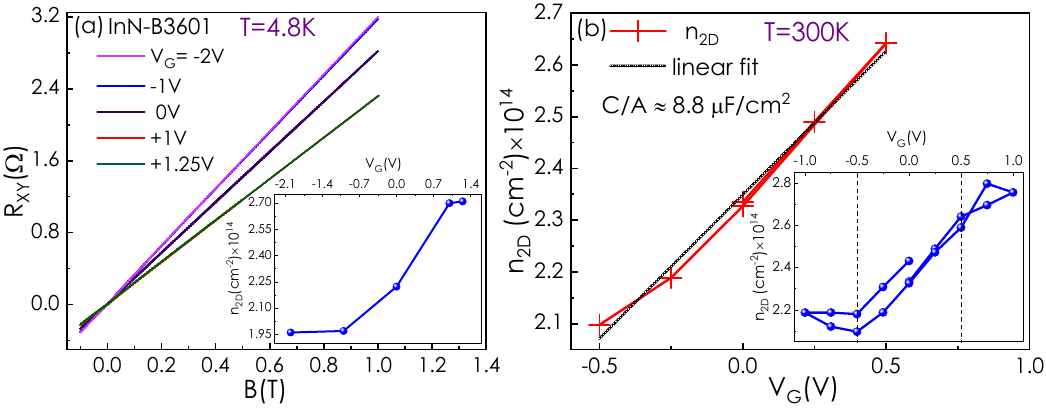}
    \caption{\textbf{Hall measurements}: (a) Hall voltage data at T=4.8 K, Carrier density modulation is shown in the inset. (b) The measured capacitance from the data over a range of gate voltages ($V_{G}$) spanning from $\pm 0.5 \rm \ V$.}
	\label{Hall_meas}	
\end{figure*}

At a temperature of $300{\rm \ {K}}$, we observe $\mu_e\sim$ $220 \rm \ cm^{2}V^{-1}s^{-1}$ at $V_{G}$ = 0 V, yielding a mean free path $l\approx{10 \rm \ {nm}}$ (see Table \ref{Summary_of_B36data}).  The capacitance determined from our measurements, approximately 8.8$\rm \ \mu Fcm^{-2}$ as depicted in Fig. \ref{Hall_meas}(b), surpasses what is achievable in MOSFET or HEMT-type devices. All low temperature measurements were performed below the glass transition temperature ($T_{G}$) of the ionic liquid (refer to supplementary data for the behavior of an ionic liquid gate). It is crucial to note that changing the gate voltage will have the desired effect only if done at $T \gg T_G$. In our experiments, we always allowed the sample to warm up to room temperature for setting $V_G$.


\begin{table*}[htbp]
	\centering
        \caption{Bulk carrier density($n_{3D}$), the density of states (DOS) at the Fermi level ($N(0)$), superconducting transition temperature ($T_{c}$), Fermi velocity ($v_{f}$), mean-free path ($l$), Ioffe-Regel parameter ($k_{f}l$), electron-phonon coupling strength ($\lambda$), coherence length ($\xi$), and Maki parameter ($\alpha$) are obtained from two devices, namely InN-B3601 and InN-B3603 respectively.}
        \label{Summary_of_B36data}
	\begin{ruledtabular}
	\begin{tabular}{ c c c c c c c c c c c  }
	\addlinespace			
        $ \rm B3601$ &   $V_{G}(\rm V)$ & $n_{3D}$$(m^{-3})$  & 
         $N(0)$($m^{-3}/eV$) & $T_{C}\rm (K)$ & $v_{f}(m/s)$  & $l(nm)$ & $k_{f}l$  & $\lambda$ &$\xi(nm)$ &$\alpha$ \\ 
              \addlinespace
              \hline 
              \addlinespace
		 &-2 & 6.53$\times$$10^{24}$ & 7.69$\times$$10^{26}$ & 2.177 & 6.68$\times$$10^{5}$  & 9.07 &  5.24 &  0.44229 & 38.82 & 0.0767 \\
			
		 &-1 & 6.56$\times$$10^{24}$ & 7.71$\times$$10^{26}$ & 2.174 & 6.69$\times$$10^{5}$  & 9.30 &  5.38 &  0.44215 &39.09 &0.0745 \\
			
		 &0 & 7.41$\times$$10^{24}$ & 8.02$\times$$10^{26}$ & 2.146 & 6.97$\times$$10^{5}$ & 9.84 & 5.93 &  0.44090 & 40.73 & 0.0699 \\
			
		 &+1 & 9.00$\times$$10^{24}$ & 8.56$\times$$10^{26}$ & 2.087  & 7.44$\times$$10^{5}$ & 9.52 & 6.12 &  0.43844 & 42.76 & 0.0649 \\
			
	   &+1.25 & 9.03$\times$$10^{24}$ & 8.57$\times$$10^{26}$ & 2.072 & 7.45$\times$$10^{5}$ & 9.35 &  6.02 & 0.43779 & 43.06 & 0.0647 \\

          \addlinespace
          \hline
          \hline
        \addlinespace
      $ \rm B3603$  & $V_{G}(\rm V)$ &  $n_{3D}$$(m^{-3})$  & 
          $N(0)$($m^{-3}/eV$) & $T_{C}\rm (K)$ & $v_{f}(m/s)$  & $l(nm)$ & $k_{f}l$  & $\lambda$ &$\xi(nm)$ & $\alpha$ \\
          \addlinespace
          \hline
          \addlinespace
          & -3.25 &  5.11$\times$$10^{24}$ & 7.09$\times$$10^{26}$  & 2.074 & 6.15$\times$$10^{5}$& 6.15 & 3.27 & 0.43788 & 31.8 & 0.1214  \\

          &  0  &   6.82$\times$$10^{24}$ & 7.80$\times$$10^{26}$  & 2.012 & 6.78$\times$$10^{5}$& 8.60 & 5.03 & 0.43519 & 37.2 & 0.0949  \\

         & +0.5 & 7.07$\times$$10^{24}$ & 7.90$\times$$10^{26}$  & 2.004 &
         6.86$\times$$10^{5}$  & 8.26 & 4.90& 0.43484& 37.4 & 0.0937  \\

         & +3.25 &   &  & 1.870 &   & &  & 0.42887  &  &  \\
         
	\end{tabular}
	\end{ruledtabular}
	
\end{table*}

\begin{figure}[h!]
	\centering

 \includegraphics[width=0.45\textwidth]{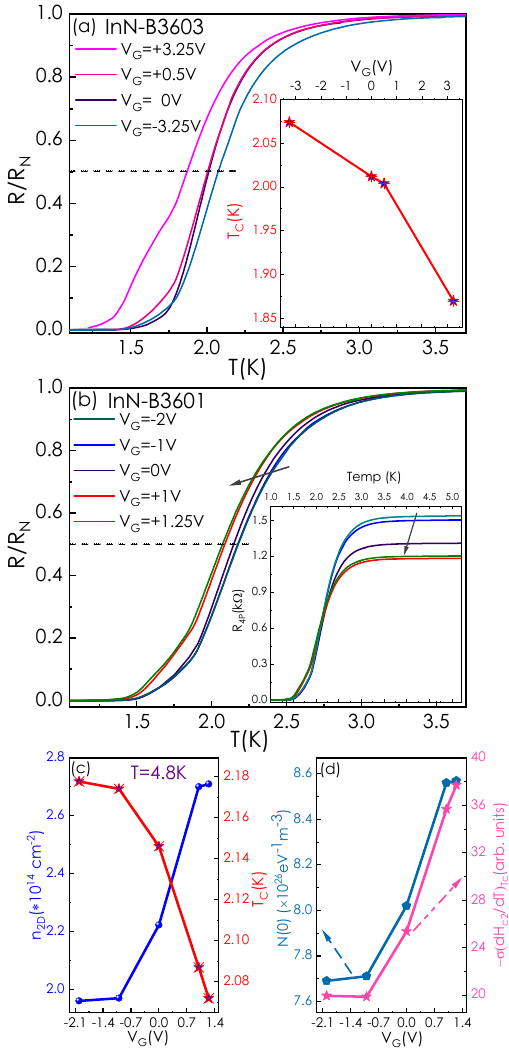}
	\caption{\textbf{Critical temperature (Tc) measurements:} (a) The normalized $R-T$ for different applied gate voltages for InN-B3603 sample. The inset displays the corresponding superconducting $T_C$ measured at various gate voltages, revealing an estimated value of $\Delta T_C$ = $204 \rm \ mK$ within the voltage range of $\pm$ $3.25 \rm \ V$. The dashed line (at 0.5) is the reference where resistance becomes half of the normal resistance ($R_N/2$). (b) Normalized $R-T$ for different gate voltages for InN-B3601 sample. The data shown in the inset are not normalized. (c) Carrier density and corresponding superconducting transition temperature for different gate voltages. Carrier density measured at $\rm T=4.8 \rm \ K$ (above the superconducting transition temperature). (d) Estimation of density of states (DOS) from the free electron theory relation (left axis) and the dirty limit relation (right axis).}
	\label{Low_temp_SC_measurement}	
\end{figure}

The superconducting critical temperature was measured in InN devices subjected to ionic liquid gating. The findings presented here stem from two such devices i.e. B3601 and B3603 (from two different wafers), demonstrating tunable superconducting parameters with variations in $T_C$ of $204$ ${\rm mK}$ and $105$ ${\rm mK}$ shown in Fig. \ref{Low_temp_SC_measurement} (a) \& (b) respectively.
The transition temperature of InN is primarily governed by carrier density rather than any disorder scattering ($k_{F}l$) \cite{dubi2007nature}\cite{ghosal1998role}. It has to be noted that we are able to both enhance and supress the $T_C$ in the same device. This cannot be explained by assuming a thin perturbed layer near the surface and a relatively unperturbed bulk superconductor in parallel. In such a case the resistive $T_C$  can be enhanced by a surface layer with a higher $T_C$ ``short circuiting'' the bulk, but cannot be suppressed by a layer with a lower $T_C$, since the bulk superconductor would then act as a shorting path. Furthermore we see no evidence of a two step resistive transition that would indicate the presence of two parallel conducting layers. The perturbation of the superconducting state that we see thus is likely to be a state spanning the entire superconducting region of the ${\rm InN-GaN}$ stack.

We find reasonable agreement  of the measured variation of the superconducting parameters with the expectations 
from a s-wave BCS superconductor having a simple free electron (density $n$) Fermi surface. The zero temperature coherence length for BCS s-wave superconductor is \cite{tinkham2004introduction}, 
\begin{equation}
  \xi(0)\approx\dfrac{\hbar v_{f}}{\pi\Delta(0)} 
\end{equation}
where, $v_f = {\dfrac{\hbar}{m}}\left( 3{\pi^2}n\right)^{1/3}$ is the Fermi velocity and $\Delta(0) \propto  T_c$ is the superconducting energy gap at $T=0$, should vary as
\begin{equation}
\dfrac{\delta\xi}{\xi(0)} \approx \dfrac{1}{3} \dfrac{{\delta}n}{n}-\dfrac{{\delta}T_c}{T_c}
\end{equation}

\begin{figure}[h]
    \centering
    \includegraphics[width=0.48\textwidth]{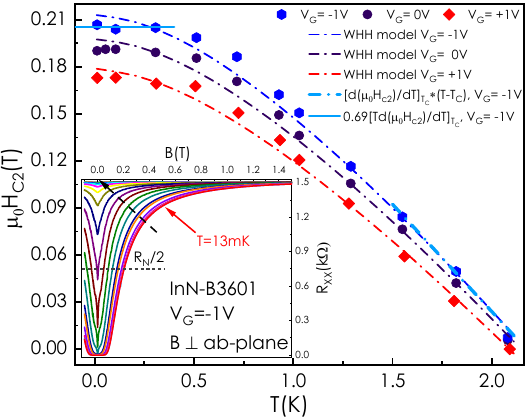}
    \caption{\textbf{Critical field ($H_{C2}$) measurements:} Temperature dependence of critical field under various applied gate voltages, fitted with WHH model. WHH model stands for Werthammer-Helfand - Hohenberg model.}
    \label{upper_critical_measurement}
\end{figure}

\begin{figure*}[t]
	\centering

     \includegraphics[width=1.0\textwidth] 
    {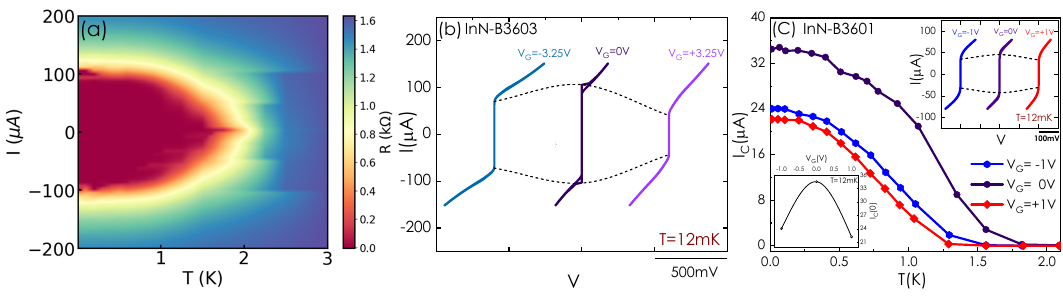}
     \caption{\textbf{Critical current ($I_C$) measurements: }(a) Critical current measurements for $V_{G}=0 \rm \ V$ for an InN hall bar. (b) The measured values of the critical current ($I_{C}$) exhibit suppression for both positive and negative applied gate voltages for InN-B3603 sample. (c) Temperature-dependent measurements of critical current ($I_{C}$) for InN-B3601 sample exhibits bipolar suppression of $I_{C}$ as shown in the inset.}
    \label{Ic_measurement}	
\end{figure*}

The value of $\xi(0)$ can be conveniently calculated from the critical field data (Fig. \ref{upper_critical_measurement}), the carrier density and $T_C$ are directly measured. This relation thus gives a good check on the consistency of the measurements, that does not require the exact constant of proportionality between $\Delta(0)$ and $T_C$ to be known. The calculated and measured values of $\dfrac{\delta{\xi}}{\xi(0)}$ differ by about $40\%$ which can be attributed to the disorder correction arising from a relatively small mean free path $l\approx 10 \rm \ nm$, that forces the observed $\xi$ to reduce to  $\xi_{obs}\approx 0.85\sqrt{{\xi(0)}l}$ \cite{de2018superconductivity} [Chap 6]. Given that we have $l\sim{10 {\rm \ {nm}}} \ll \xi(0)$ a correction of the order of what we observe is certainly expected. However given that ${k_f}l \gg 1 $,  we do not expect any strong localisation effects or significant effects of granularity. This is also supported by hysteresis-free ${\rm IV}$ characteristics in the critical current measurements as shown in Fig. \ref{Ic_measurement}.

We thus consider free electron estimation of the density of states of at the Fermi level 
$N(0)=\dfrac{m}{{\pi^2}\hbar^2}\left( 3{\pi^2}n \right)^{1/3}$ to be  reasonable. This can be used to infer the variation of the strength of the attractive interaction $V$, near the Fermi surface via the determination of $\lambda = N(0)V$ by inverting the Eliashberg-McMillan eq.,

\begin{equation}
T_C=\dfrac{\Theta}{1.45} \cdot \rm exp\left( \dfrac{-1.04(1+\lambda)  }{\lambda-\mu^*(1+0.62\lambda)}        \right)
\end{equation}
where $\mu^{*} \approx 0.1$ is the Coulomb pseudopotential \cite{mcmillan1968transition}, which accounts for the electron-electron repulsive interaction and $\Theta$ $\approx$ 370 K is the Debye temperature \cite{davydov1999experimental}. The coupling constant found here ($\lambda$$<$1) suggests that $\rm InN$ behaves like a weak to medium coupling superconductor. Clearly, at lower carrier densities the pairing interaction strength gets stronger resulting in a higher $T_C$ as depicted in Fig. \ref{Low_temp_SC_measurement}(c). The zero temperature coherence lengths $\xi(0)$ calculated from the extrapolated values of the critical field ( Table-\ref{Summary_of_B36data}), also mirror this increase of $T_C$ showing the superconducting state is stronger at lower carrier densities.

Furthermore, it is known that DOS is proportional to the slope of the upper critical field \cite{werthamer1966temperature,guo2004superconductivity, chockalingam2008superconducting}, i.e.
 \begin{equation}
N(0)\propto -{\sigma} \left( \dfrac{dH_{c2}}{dT} \right)_{T_C}
\end{equation}
The calculated $N(0)$ is shown in the right panel of Fig. \ref{Low_temp_SC_measurement}(d) and compared with the free electron values $N(0)=\dfrac{m}{\pi^2\hbar^2}\left(  3{\pi^2}n\right)^{1/3}$ exhibiting a very good agreement.

Next we looked into the temperature dependence of critical field. The magnetic field value (normal to the surface) at which the resistance of the superconductor attains half of its normal state value (the mean field value) is taken as critical field ( $H_{C2}$). $H_{C2}(T)$ curves were fitted using the well-known Werthammer-Helfand-Hohenberg (WHH) model \cite{werthamer1966temperature, helfand1966temperature} to understand the role of spin-orbit coupling (SOC), if present in the system, on the superconducting phase. WHH model gives the eq. \ref{WHH_eq}.

\begin{align}
     \ln \left ( {\dfrac{1}{t}} \right )&=\sum_{\nu=-\infty}^{\infty} \left ( \dfrac{1}{|2\nu+1|} - \left [|2\nu+1| \right. \right. \nonumber \\ 
     & \left. \left. +\dfrac{b_{c}}{t}+\dfrac{(\alpha b_{c}/t)^2}{|2\nu+1|+(b_{c}+\lambda_{so})/t} \right ]^{-1} \right )
     \label{WHH_eq}     
\end{align}

where, $t=T/T_{C}$, $\alpha =3/2m{v_F}^2 \tau$, is the Maki parameter, the dimensionless magnetic field $b_{c}$=$\mu_0 h_c$=$\dfrac{e\hbar}{2m\pi\alpha} \dfrac{\mu_0 H_{c2}}{k_B T_C}$ and $\lambda_{so}$ is the spin-orbit scattering constant. The analysis reveals a small Maki parameter $\alpha < 0.1$ and virtually no impact of the spin-orbit term within the measured range of carrier densities (see Table \ref{Summary_of_B36data}). That is further confirmed by evaluating the limiting orbital field, $H_{c2}^{orb}$, using the expression $H_{c2}^{orb}$ (T=0) $\approx$ $0.7 \lvert T \frac{dH_{c2}}{dT} \rvert_{T_C}$. The calculated values closely correspond to the experimental data, as demonstrated for a gate voltage of $V_G = -1 \rm \ V$ (sky blue line) in the Fig. \ref{upper_critical_measurement}. In cases where spin-orbit coupling is strong, one obtains a high value of Maki parameter ($\sim$ 2-3), as observed in our earlier work \cite{chakraborti2022formation}.

However there is one significant aspect in which an indicator of the strength of the superconducting state follows an opposing trend. In general the Landau-Ginzburg calculation of the critical current ($I_C$) of a thin film \cite{de2018superconductivity} [Chap 6] would predict a variation $I_C(T) \propto \dfrac{1}{\xi(T)}$ and hence approximately increase with $\sqrt{H_{C2}}$. However we find that the application of a gate voltage either positive or negative substantially suppresses the critical current by as much as $60\%$ (see Fig. \ref{Ic_measurement}(b)). Many recent experiments demonstrated gate controlled critical current suppression in superconducting films  \cite{rocci2020gate,de2018metallic,paolucci2021electrostatic,golokolenov2021origin,alegria2021high}. A set of experiments suggest that critical current suppression is occurring due to Cooper pair breaking due to gate field \cite{de2018metallic,rocci2020gate,paolucci2021electrostatic}. Most likely the effect is due to modification of the electron distribution in the channel in response to the additional electric field arising from the gate. It is possible that the wavefunction as it either gets pulled towards the surface or as it gets pushed into the non-superconducting bulk region in fact sees regions with smaller normal-state mean free path and subsequent surface scattering and granularity effects leading to the suppression of the critical current.

The ability to change carrier density in a material to such an extent that the Fermi surface runs through a large volume of the Brillouin zone, potentially changing its shape from simple spherical or elliptical to more complex geometries incorporating nodes, necks, nesting vectors etc. does not exist as of now. For example in a simple tight binding on a square lattice, the Fermi surface changes from a simple circle to a square at half filling and then opens up. Development of techniques and materials in which  potentially very large density modulations are possible are  important as electron interaction driven phenomena inherently depend on the shape of the Fermi surface. In this paper, we have shown an example that can be a  step forward in this direction.\\

Acknowledgements: We acknowledge support from Department of Science and Technology, Government of India under project: RD/0120-DSTIC01-001, Science and Engineering Research Board (SERB), Government of India under project: CRG/2018/001405. BCB acknowledges financial support through DST-INSPIRE fellowship of Govt. of India. We thank Amit P Shah (TIFR, Mumbai) for his assistance with the dry etch process and Prof. Saibal
Sarkar (Department of Energy Science and Engineering, IIT Bombay) for providing access to deposit $\rm Al_2 O_3$ films. We thank Dr. Parushottam Majhi for useful discussions.

\bibliographystyle{apsrev4-2}
\bibliography{references}

\onecolumngrid
\clearpage
\mbox{}


\section*{Supplementary material}

# Extended data
# Gate voltage modulation of the superconducting state in a degenerate semiconductor

Bikash C. Barik,[1] Himadri Chakraborti,[1,2] Buddhadeb Pal,[1,3] Aditya K. Jain,[1,4] Swagata Bhunia,[1] Sounak Samanta,[1] Apurba Laha,[5] Suddhasatta Mahapatra,[1] and K. Das Gupta[1,*]

[1]*Department of Physics, Indian Institute of Technology Bombay, India*
[2]*SPEC, CEA, CNRS, Universite Paris-Saclay, CEA Saclay, 91191 Gif sur Yvette Cedex France*
[3]*Indian Association for the Cultivation of Science, Kolkata, 700032, India*
[4]*Department of Physics, Royal Holloway, University of London, Egham, Surrey, TW20 0EX, UK*
[5]*Department of EE, Indian Institute of Technology Bombay, India, India*

## I. INDIUM NITRIDE SAMPLES FABRICATION PROCESS

Here we give a detailed idea about the growth technique, fabrication processes and different characterization techniques of InN samples that we have used in our experiment. InN samples were grown in plasma-assisted molecular beam epitaxy (PA-MBE) system (RIBER compact 21) with RIBER Addon radio frequency plasma source for the active N species on sapphire substrates and this growth was done by Kankat Ghosh/Swagata Bhunia [1]. First GaN layer of ∼300 nm is grown on the sapphire substrate and then ∼300 nm InN is grown on the top of that. To make the channel of InN, etching needs to be done. The sample goes through different stages during fabrication which are mentioned below.

### Fabrication steps

All the fabrication steps are described below.

- Substrate preparation: Cleaning of the sample with acetone and isopropyl alcohol.

- Deposition of 100nm $Al_2O_3$ as a hard mask using atomic layer deposition (ALD).

- Hall bar pattern by photolithography. It includes:
    - Spin coating on photoresist using S1813 positive photoresist.
    - Exposing to UV light through a mask using mask aligner (MA400).
    - Development in the MF319 developer solution.

- Etching $Al_2O_3$ in HF solution.

- Dry etch of InN in reactive etching system (RIE).

- Removal of photoresist and $Al_2O_3$.

- Patterning contact by photolithography.

- Metals (Ti/Au) deposition by thermal evaporation process of thickness 20nm and 100nm.

- Liftoff in acetone.

- Mounting in a Leadless chip carrier (LCC) and bonding is done by a wire bonder.

- Deposition of ionic liquid on the hall bar and curing for 48 hours at high vacuum (Pressure $< 1 \times 10^{-5}$ mbar) and at temperature 110°C.

---

* kdasgupta@phy.iitb.ac.in

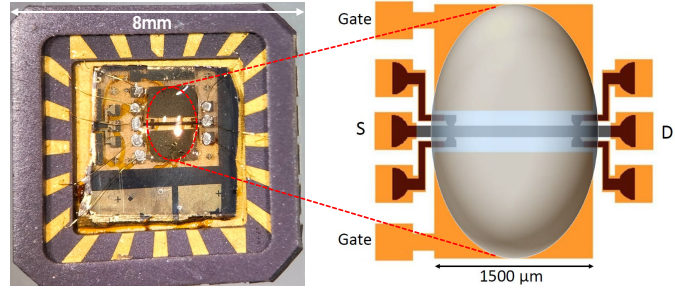

FIG. 1. Actual InN device, ionic liquid deposited portion is shown in the red-dashed line (left). Schematic diagram of the device is shown. (right)

## II. TRANSISTOR CHARATERSTICS

Fig. 2 shows the transfer characteristics of the IL-gated InN at room temperature. To maintain a constant source-drain voltage of 100 mV ($V_{SD}$), a DC voltage source is used. The gate voltage ($V_G$) is swept in 0.1 V steps using another voltage Source. Notably, different settling times—ranging from 30 s to 240 s are set during the gate voltage sweep. These settling times allow for the system to stabilize after voltage changes. The measurement of gate leakage current is crucially smaller by approximately three orders of magnitude than the source-drain current. This substantial difference ensures that gate leakage doesn't impact on the device's performance. The observed trend in the data indicates $n$-type characteristics, as $I_{SD}$ increases with the application of increasingly positive $V_G$.

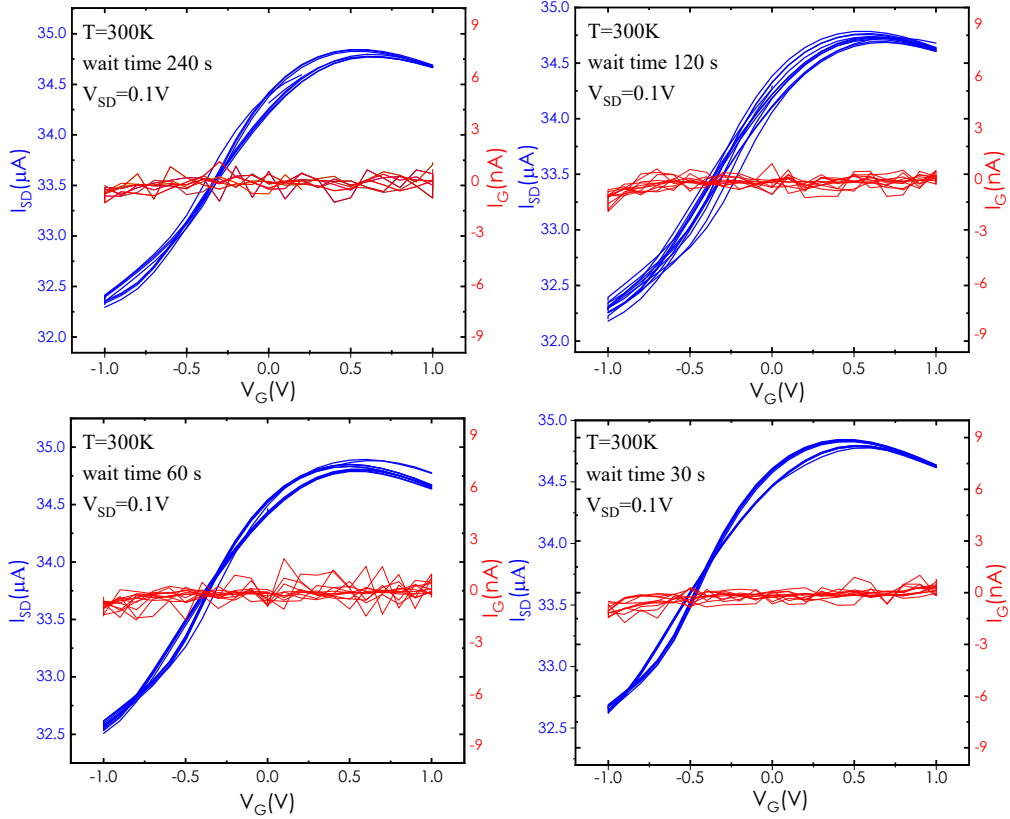

FIG. 2. At T=300 K, the transfer characteristics exhibit a nearly hysteresis-free nature across various settling times (InNB3601 sample).

## III. BEHAVIOUR OF AN IONIC LIQUID GATE

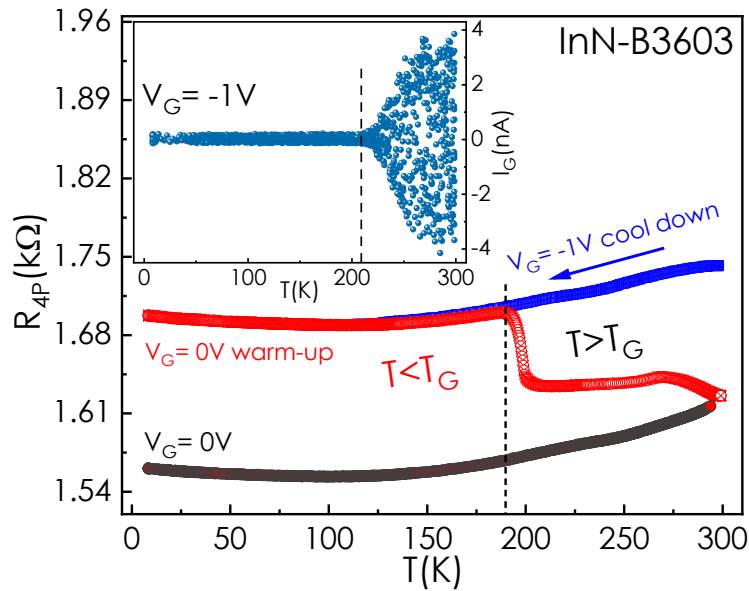

FIG. 3. The glass transition temperature phenomenon is investigated by applying a gate voltage of $V_G = -1$ V and subsequently removing the gate voltage at a low temperature of 10 K. The resulting observation of a significant jump in resistance serves as a clear indicator of the glass transition temperature associated with the ionic liquid. The inset depicts the recorded gate leakage current at a gate voltage, $V_G = -1$ V, providing a compelling illustration of the phenomenon associated with the freezing of ions.

Fig 3 shows the striking difference between the behaviour of an ionic liquid gate and a conventional dielectric gate. The ionic motion within the liquid starts to undergo a slowdown below approximately $\sim 230$ K. Furthermore, below a specific temperature marked as the glass transition temperature ($T_G$), the ions associated with the ionic liquid experience complete immobilization. [2][3]. A gate voltage $V_G = -1$ V was applied and the sample was cooled down to 10 K. The resistance follows the blue curve. The corresponding gate leakage current is illustrated in the inset. At 10 K the voltage source is set to zero and disconnected. The sample is then allowed to warm up but continues to behave as if the gate voltage is still present. When the glass transition temperature ($T_G$) is reached, the ions in the ionic liquid regain their motion and adjust to the absence of the external field, at this point one begins to see considerable fluctuations in the gate leakage current. This happens over a range of $T \approx 180-200$ K which corresponds well to the known value of $T_G$ of DEME-TFSI. Changing the gate voltage will have the desired effect only if done at $T \gg T_G$. In our experiments we always allowed the sample to warm up to room temperature for setting $V_G$.

## IV. CAPACITANCE MEASUREMENTS

### A. From gate response

The time response of the electrical double layer (EDL) gate in the InN film to a step change in gate bias voltage ($V_G$) has been investigated. The results are illustrated in Fig. 4 depicting the growth and decay curves for various gate voltages.

The figures showcase the InN film resistance's time response to negative gate voltages, revealing distinct growth and decay curves. The time scale for capacitor charging and discharging is extracted from these curves. Notably, for $V_G$= -0.5 V, the fast time scale ($\tau$) is found to be consistent for both growth (6.9 s) and decay (9.5 s) processes, indicating the charging and discharging of the EDL. The gate leakage current suggests that gate charging occurs through an equivalent resistance ($R_G$) of approximately $\simeq 700$ MΩ.

The time constant of the charging and discharging curves ($\tau = RC$) is utilized to determine the gate capacitance ($C_G$). For $V_G$= -0.5 V, the calculated gate capacitance is approximately $\sim 8.52$ $\mu$F/cm$^2$. A similar analysis for $V_G$ = -1 V yields a gate capacitance of around $\sim 10.4$ $\mu$F/cm$^2$, with gate charging occurring through an equivalent

resistance of $R_G \simeq 400$ MΩ, as depicted in figure 4. The gate capacitance values measured by this technique match well with the capacitance derived from Hall measurements i.e 8.8 $\mu$F/cm$^2$ (see text).

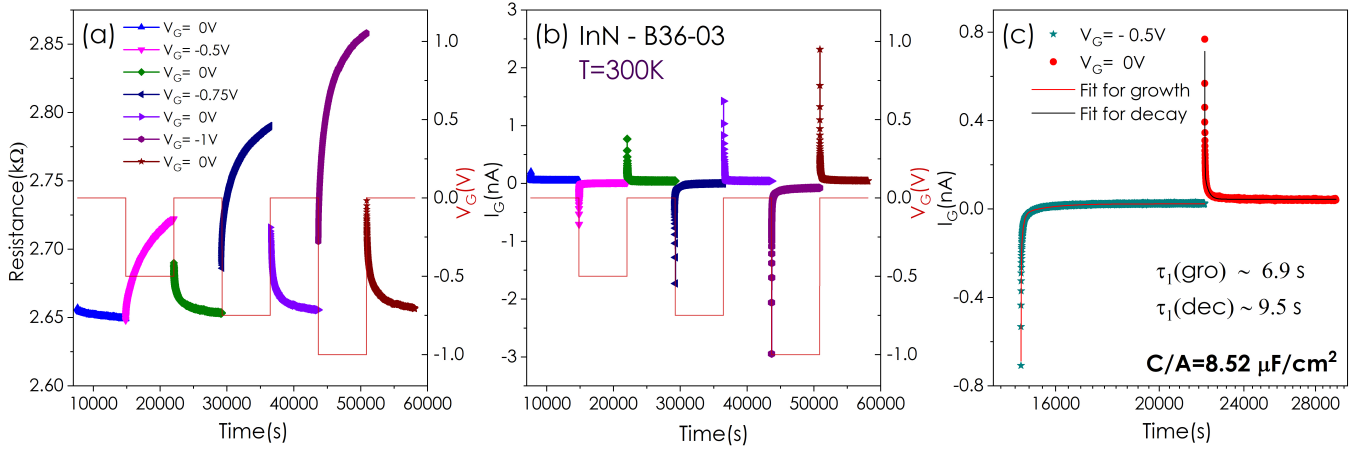

FIG. 4. (a) Transient response of InN film resistance to the negative gate voltages. (b) The gate leakage current of corresponding gate voltages are shown. (c) Transient response of the gate leakage current for $V_G$= -0.5 V. The calculated value of the capacitance from the fitting is shown.

### B. From electrochemical impedance analysis

We used electrochemical impedance spectroscopy to study the frequency dependence impedance as well as to measure the EDL capacitance [4][5]. We measured the impedance of gated InN thin film as a function of frequency for different applied gate voltages from 37 mHz to 2 kHz. Impedance measurements are performed using two lock-in amplifiers. The low-frequency measurement is done by configuring a high time constant of 30 s and a corresponding settling time of 300 s. The gate voltage is applied using a voltage source. Lock-in amplifiers do not give the value of impedance directly but it can be calculated using the following relations.

$$Z = \frac{V_S}{I_S} = \frac{V_S}{V_R} \times R = R \times \left( \frac{V'_S + jV''_S}{|V_R|} \right) \tag{1}$$

$$Z = R \times \left( \frac{V'_S + jV''_S}{\sqrt{(V'_R)^2 + (V''_R)^2}} \right) \tag{2}$$

$$Z' = R \times \frac{V'_S}{\sqrt{(V'_R)^2 + (V''_R)^2}} \tag{3}$$

$$Z'' = R \times \left( \frac{V''_S}{\sqrt{(V'_R)^2 + (V''_R)^2}} \right) \tag{4}$$

Where $V_S$ and $V_R$ are the voltage across the InN sample and standard 1$k\Omega$ resistor respectively. 1 kΩ is used to measure the current through the sample. $Z'$ and $Z''$ are real and imaginary parts of the impedance and $Z$ is the total impedance of the system. The model shown in the inset of Fig. 5 is the capacitative coupling between InN film with resistance $R$ and the ionic liquid with resistance $R_{IL}$. A constant current 10 $\mu$A is applied by applying 2.5 V from the lock-in's internal oscillator through a 250 kΩ resistor in series. The magnitude of the impedance between the source and drain can be calculated. Fig. 5 show the impedance of the InN sample for different applied gate voltages.

In the model shown in the inset of Fig. 5, the capacitance near the gate-liquid interface is taken to be approximately 13 times greater than the channel-liquid capacitance. This ratio is justified by the substantial difference in the areas

of the gate electrode and the channel, indicating that the majority of the potential will be distributed across the working electrode, i.e. the source and drain regions. Consequently, the need for a reference electrode is eliminated. The circuit model, as illustrated in the inset of Fig. 5, is fitted for impedance analysis. Notably, a satisfactory fit is achieved using this model up to 1 kHz, and the resulting values for capacitance per unit area and $R_{IL}$ are detailed in Table I.

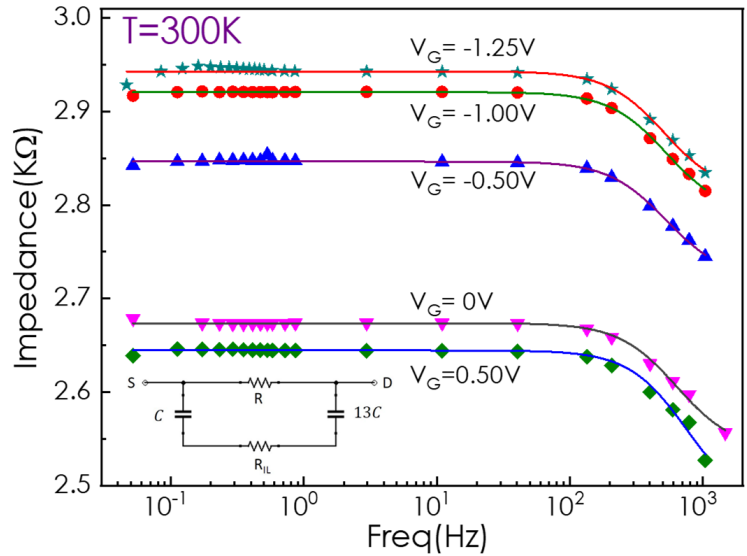

FIG. 5. (a) Frequency dependence impedance of InN thin film at different mentioned gate voltages. Solid lines are the fits using circuit models shown in the inset of Figure. $R$, $R_{IL}$ are the resistances of the sample and ionic liquid respectively.

TABLE I. Capacitance and resistances of ionic liquid for different applied gate voltages obtained from EIS.

| $V_G$ | Capacitance($\mu F/cm^2$) | $R_{IL}$(k$\Omega$) |
|---|---|---|
| 0.5 | 4.84 | 38.6 |
| 0 | 4.5 | 50 |
| -0.5 | 4.4 | 61.8 |
| -1 | 4.33 | 61.9 |
| -1.25 | 4.45 | 62.6 |



\end{document}